\documentclass[%
 reprint,
superscriptaddress,
 amsmath,amssymb,
 aps,
prl,
]{revtex4-2}
\usepackage{graphicx}% Include figure files
\usepackage{dcolumn}% Align table columns on decimal point
\usepackage{bm}% bold math
\usepackage{siunitx} % very nice little package for SI units
\usepackage{xcolor}
\usepackage{hyperref}% add hypertext capabilities
\newcolumntype{"}{@{\hskip\tabcolsep\vrule width 1pt\hskip\tabcolsep}}
\makeatletter
\newcommand{\thickhline}{%
    \noalign {\ifnum 0=`}\fi \hrule height 1pt
    \futurelet \reserved@a \@xhline
    }
\newcommand{\ket}[1]{\lvert#1\rangle} % Ket
\newcommand{\braket}[2]{ \langle #1 | #2 \rangle} %Inner Product
\newcommand{\R}{\mathrm{R}} % recoil subscript
\newcommand{\GKP}[1]{\lvert \mathrm{GKP} #1\rangle}

\usepackage{verbatim}
\newcommand{\detailtexcount}[1]{%
  \immediate\write18{texcount -merge -sum -q main_rewrite.tex output.bbl > main_rewrite.wcdetail }%
  \verbatiminput{main_rewrite.wcdetail}%
}

\begin{document}
\preprint{APS/123-QED}

\title{Error Correcting States in Ultracold Atoms}

\title{Error Correcting States in Ultracold Atoms}

\author{Harry C. P. Kendell}
\affiliation{Quantum Engineering Centre for Doctoral Training, University of Bristol, Bristol BS8 1FD, UK}
\affiliation{Quantum Engineering Technology Laboratories, H. H. Wills Physics Laboratory and Department of Electrical, Electronic, and Mechanical Engineering, University of Bristol, Bristol BS8 1FD, UK}

\author{Giacomo Ferranti}
\affiliation{Quantum Engineering Technology Laboratories, H. H. Wills Physics Laboratory and Department of Electrical, Electronic, and Mechanical Engineering, University of Bristol, Bristol BS8 1FD, UK}
\author{Carrie A. Weidner}
\email{c.weidner@bristol.ac.uk}
\affiliation{Quantum Engineering Technology Laboratories, H. H. Wills Physics Laboratory and Department of Electrical, Electronic, and Mechanical Engineering, University of Bristol, Bristol BS8 1FD, UK}

\noaffiliation

\date{\today}% It is always \today, today,

\begin{abstract}
We demonstrate a method for encoding Gottesman-Kitaev-Preskill (GKP) error-correcting qubits with single ultracold atoms trapped in individual sites of a deep optical lattice. Using quantum optimal control protocols, we demonstrate the generation of GKP qubit states with $\SI{10}{\deci\bel}$ squeezing, which is the current minimum allowable squeezing level for use in surface code error correction. States are encoded in the vibrational levels of the individual lattice sites and generated via phase modulation of the lattice potential. Finally, we provide a feasible experimental protocol for the realization of these states. Our protocol opens up possibilities for generating large arrays of atomic GKP states for continuous-variable quantum information.
\end{abstract}

\maketitle

\section{\label{sec:intro}Introduction} 
Fault-tolerant quantum information processing will require fast low error-rate operations on qubits~\cite{preskill}. This ensures large algorithms can be fully implemented without the exponential build up of errors rendering the final result useless. However, current systems suffer from relatively high physical error rates~\cite{Brandhofer2021}. This must, at least in the near term, be offset through the use of error-correcting codes~\cite{laflamme1997}. These rely upon a larger state space upon which a smaller subspace of information is encoded. This proceeds either through the representation of a single logical qubit with multiple physical qubits or through the use of a bosonic oscillator, which itself describes a larger state space.

The Gottesman-Kitaev-Preskill (GKP) states~\cite{GKP2001_PROPOSAL} are the canonical example of this latter case. These states have been shown to have the highest error-correction performance compared to similar encodings~\cite{albert2018_CODE_PERFORMANCE}, and they have been shown to have utility in cluster-state quantum computing~\cite{GKP_cluster}. GKP states consist of periodic delta functions evenly spaced in position space. The two encoded qubit states are identical, except that one is displaced by half the spacing period. This nominally allows for perfect error correction up to a displacement by a quarter of the inter-delta spacing. Finite approximations to these ideal infinitely-squeezed states provide consequently approximate error correction.

The difficulty in preparing these states lies in the challenge of generating adequate squeezing to offset the overhead of the encoding and reduce the logical error rate. 
Furthermore, the control of these states is very sensitive to noise, for more accurate approximations, due to the large amplitude in phase space~\cite{Grimsmo2021_REVIEW}. These states must therefore be considered not only in their error-correction ability but also the difficulty with which they may be implemented and controlled. 
Nevertheless, approximate GKP states have been experimentally demonstrated in ions~\cite{fluhmann2019_FIRST_ION_GKP}, transmon qubits~\cite{campagne2020_FIRST_TRANSMON_GKP}, and photonic systems~\cite{konno2023_FIRST_OPTICAL_GKP}.

Here, we present a proposal for the realization of GKP states in ultracold atoms, specifically, arrays of single atoms trapped in the individual sites of a deep optical lattice potential. Atoms are excellent candidates for qubits, and a reprogrammable tweezer-based quantum information processor with error correction has recently been demonstrated~\cite{lukin2023}, where quantum information is processed via atom-atom interactions mediated by selective excitation of atoms into high-lying Rydberg states~\cite{Adams_2020}. Our protocol can be straightforwardly adapted to atoms in optical tweezers~\cite{lukin_tweezers, browaeys_tweezers, regal_tweezers}, but we focus here on atoms in optical lattices. This is due to their straightforward scalability and controllability, even at the single-atom level~\cite{greiner2009,kuhr2010,kuhr2011}. Additionally, optical-lattice-based systems have found use in an array of high-precision quantum technologies, including atomic clocks~\cite{Yb_lattice_clock,Sr_lattice_clock}, quantum simulators~\cite{qsim1,qsim2}, and inertial sensors~\cite{PdS2013,SLI_exp,mueller,Rasel2021}.

\section{\label{sec:theory}Background}

The GKP states have a rich history of approximation, given the full code states are physically unrealizable. A number of these possess attractive qualities, such as the representation of approximate GKP states as sums of translated coherent states, which naturally admits an experimental construction in ions~\cite{fluhmann2019_FIRST_ION_GKP}. However, these approximations largely agree up to a redefinition of parameters~\cite{matsuura2020_GKP_EQUIVALENCE}. We therefore choose, without loss of generality, an approximation as a series of displaced squeezed vacuum states with an overall Gaussian envelope. We additionally work with symmetric GKP states in the two quadratures, motivated by the fact there is no \emph{a priori} reason to weight error correction performance to one quadrature. This introduces an additional squeezing factor $\hat{S}_0$~\cite{matsuura2020_GKP_EQUIVALENCE}.

We denote the GKP code states by $\GKP{0,1}$ and Fock states by $\ket{n}$. The finite GKP code state is, therefore, up to normalization,
\begin{equation}
\begin{split}
    \GKP{k} \propto \hat{S}_0   \sum_{s \in \mathbb{Z}} e^{-\frac{1}{2}\sigma^2x_t^2}   \hat{X}(x_t)   \hat{S}(\Delta)   \ket{0}.
\end{split}
\end{equation}
The displacement value $x_t= \sqrt{\pi}(2s+k)$ for each squeezed vacuum state, given the code state $k \in \{0,1\}$. 
The additional squeezing factor is given by
\begin{equation}
    \hat{S}_0 = \hat{S}\big(\ln(\sqrt{1+\sigma^2\Delta^2})\big),
\end{equation}
with $\Delta = -\ln\big[\sqrt{\sigma^2/(1-\sigma^4)}\big]$. 
The finite GKP state is then characterised solely by the level of squeezing as $\zeta = -10\log_{10}(\sigma^2)  \SI{}{\deci\bel}$.

Here, we find optimal controls that can be used to generate GKP states with $\zeta = \SI{10}{\deci\bel}$ of squeezing with atoms in optical lattices. This level of squeezing is generally recognized to be the point at which such states are useful from a quantum information perspective~\cite{andersen,furusawa}. These states may also be concatenated with a surface code~\cite{surface_review}. This provides enhanced error correction performance at the cost of additional overhead and complexity. As a result, fault-tolerant quantum computing can be achieved at as low as $\SI{8.1}{\deci\bel}$ of squeezing \cite{fukui2023_8_1dB}, assuming zero losses in the system.

The Hamiltonian of the optical lattice system used to generate these states takes the form
\begin{equation}
    \label{eq:H}
    H = \frac{p^2}{2m} + U\cos{\big(2k_\mathrm{L}[x + u(t)]\big)}
\end{equation}
where the lattice wavenumber is $k_\mathrm{L} = 2\pi/\lambda$ for a lattice of wavelength $\lambda$, the lattice depth is $U$ (typically quoted in units of the photon recoil energy $E_\R = \hbar^2k_\mathrm{L}^2/2m$ for an atom of mass $m$), and $u(t)$ is our control, with units of length. By varying $u(t)$, the lattice is phase modulated, driving transitions between different external atomic states while maintaining the atom's internal state.

\begin{figure}[ht!]
    \centering
    \includegraphics[width=\columnwidth]{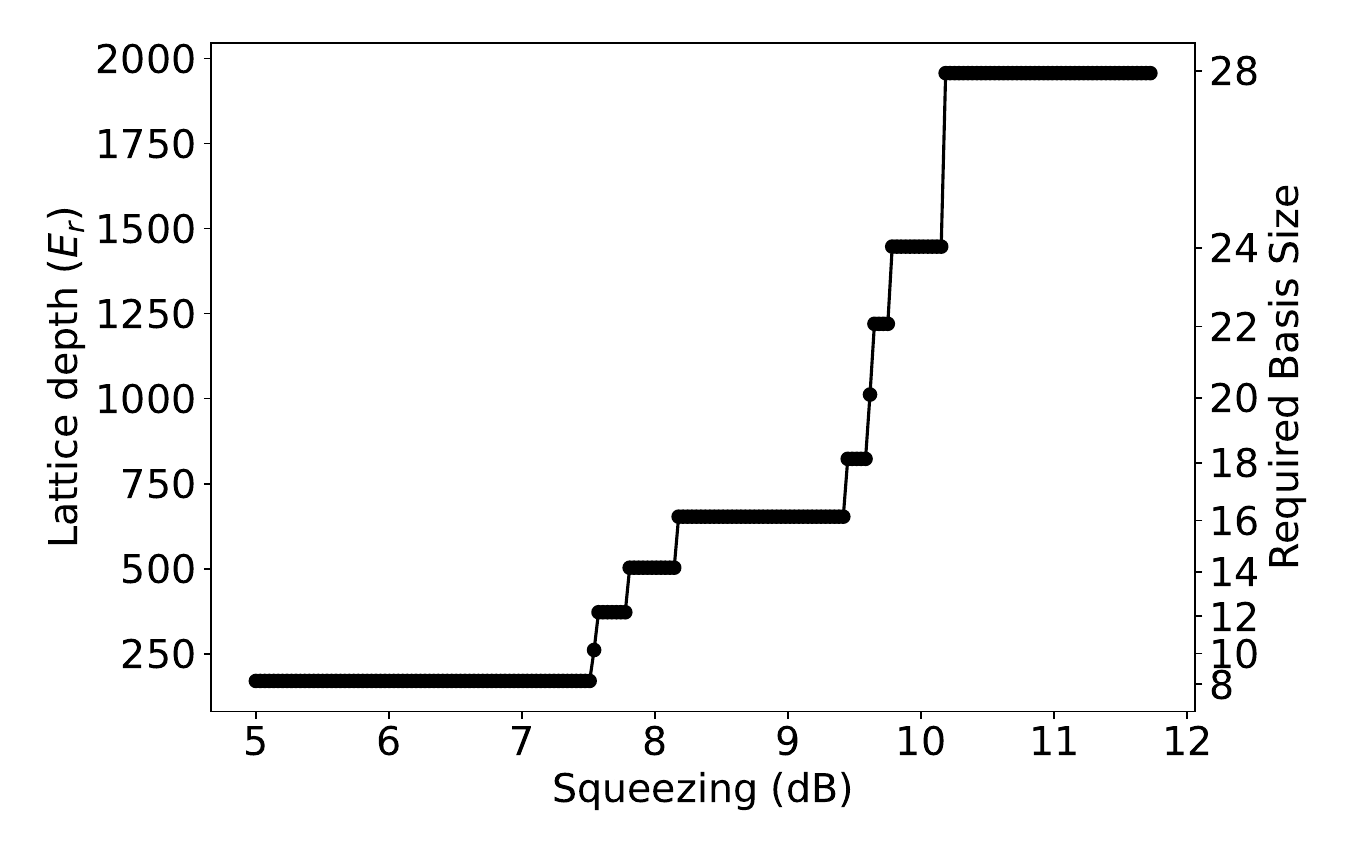}
    \caption{Lattice depth (left axis) and required basis size (right axis) as a function of desired squeezing level where the basis size is chosen such that the GKP state can be reconstructed with a fidelity of $\mathcal{F}\geq 0.99$. The large plateaus in depth and basis size arise due to the fact that, as more accessible Fock states are introduced into the problem, there is a range of available squeezing parameters realizable to within our target fidelity.}
    \label{fig:sqz-basis}
\end{figure}

We assume the atom is trapped in a single site of a deep 1D lattice potential, which is straightforward to experimentally prepare~\cite{Greiner2002, kuhr2010}. We then define, as in Ref.~\cite{Winkelmann2022}, our Fock states $\ket{n}$ to correspond to the bound vibrational levels within a given lattice site, i.e. $n = 0, ..., N$. Thus, the total number $N$ of Fock states bound to a single lattice site directly affects the maximum obtainable squeezing of our resulting GKP states. Figure~\ref{fig:sqz-basis} shows, for a given squeezing level, the required lattice depth and resulting Fock basis size (i.e., the number of vibrational states bound to a given lattice site). The purpose of our phase modulation is then to control transfers between these Fock states to generate and manipulate the desired GKP states.

In our simulations, we initialize our atomic state in a given initial state $\ket{\psi(t=0)}$ (here, $\ket{0}$, considering the generation of GKP states from the ground state of the lattice potential). We then generate an initial guess (seed) for our shaking function $u(t)$ where $0\leq t \leq T$ and use the symmetrized split-step method~\cite{splitstep} to evolve the state using the time-dependent Schr{\"o}dinger equation. This results in a final state $\ket{\psi(t = T)} = \ket{\psi_\mathrm{F}}$ where the time $T$ is varied; higher values of $T$ typically lead to higher fidelity protocols. The optimization protocol then adjusts $u(t)$ in an attempt to maximize fidelity.

We define our state transfer fidelity as $\mathcal{F} = |\braket{\psi_\mathrm{D}}{\psi_\mathrm{F}}|^2$, 
where $\ket{\psi_{\mathrm{D,F}}}$ are the desired state and final state after a given shaking protocol $u(t)$, respectively. In Fig.~\ref{fig:sqz-basis} and this work in general, our target fidelity is $\mathcal{F}\geq 0.99$; extensions beyond this will require more bound states and a deeper lattice, posing increasing experimental challenges. It remains to be seen what fidelities are required for atomic GKP states to be useful in quantum information protocols; finite fidelity likely manifests similarly to loss in photonic systems. 

To find our controls, we use the GRAPE method~\cite{grape} with BFGS built into the QEngine C++ library~\cite{sorensen2019_QENGINE}. 
Furthermore, bandwidth and amplitude limits for $u(t)$ are implemented via regularization and soft boundary terms in our optimization functional. 
Practically, we used a smooth low pass filter centered at $\SI{0.5}{\mega\hertz}$ to ensure the experimental viability of the controls~\cite{Dupont2021}, in addition to limiting the shaking amplitude to no more than half a lattice site.

\section{\label{sec:exp}Generating GKP states in atoms}

\begin{figure*}[ht!]
    \centering
    \includegraphics[width=\textwidth]{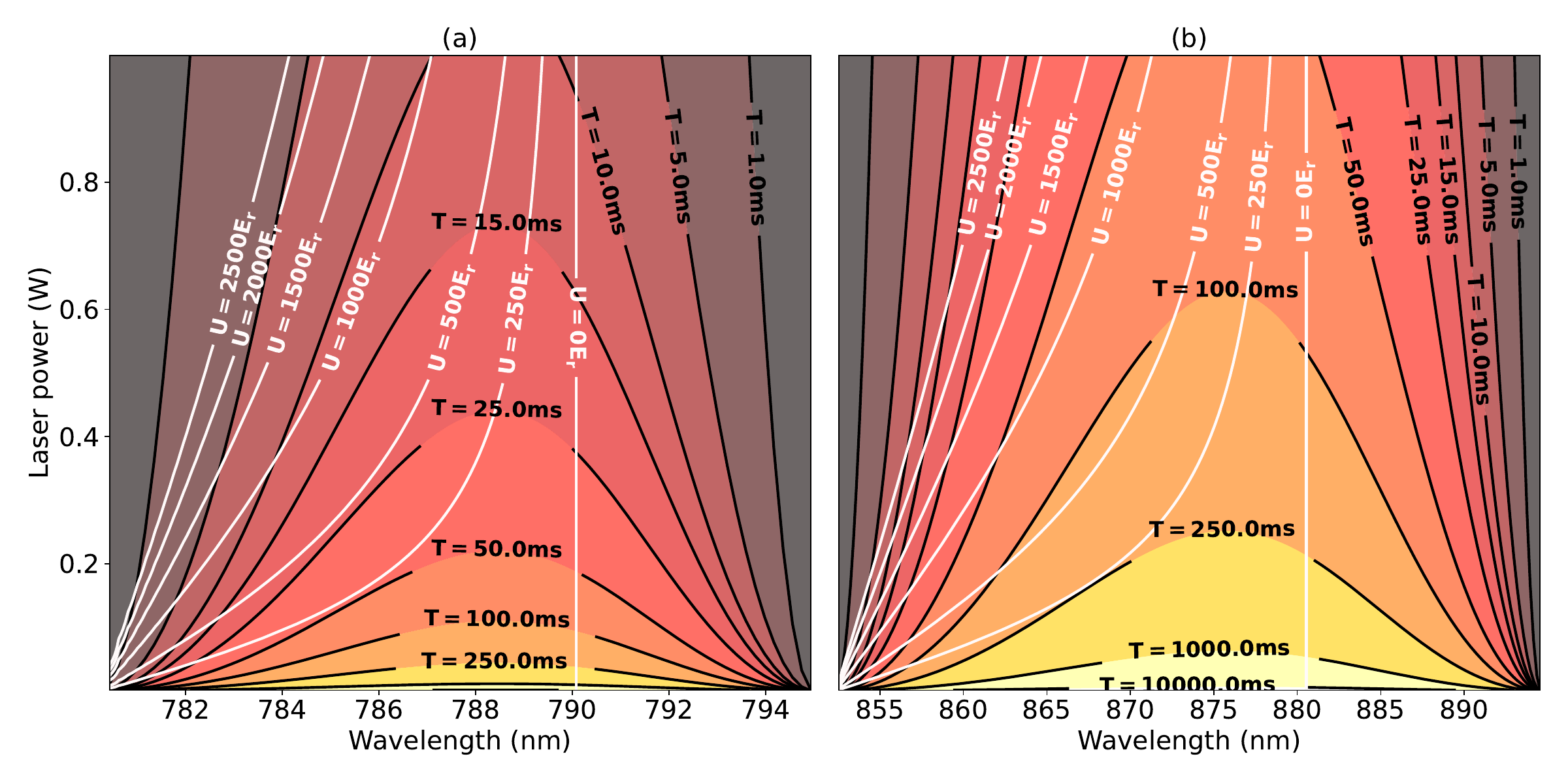}
    \caption{Contour plot of lattice trap lifetime $T$ (black lines and shading) and lattice depth $U$ (white lines) as a function of laser power and wavelength between the D1 and D2 lines for (a) rubidium and (b) cesium, assuming a 1D lattice with a beam waist of $\SI{150}{\micro\meter}$. For the lattice depths of $U\approx 1500 E_\R$ that give rise to useful squeezing levels of $\approx\SI{10}{\deci\bel}$, the cesium atom lifetime is roughly a factor of six longer than that of rubidium.}
    \label{fig:depth-lifetime}
\end{figure*}

The experimental proposal requires atoms to be loaded into the ground state of an optical lattice generated by two lasers with identical wavelength but varying phase. Ground-state loading can be achieved by either loading from a Bose-Einstein condensate or sideband cooling~\cite{Jessen1998,Lan2021} atoms loaded into the lattice directly after laser cooling. Phase modulation can be achieved by changing the phase of one lattice beam relative to the other, e.g., by controlling the RF tone sent to an acousto-optic modulator (AOM). In this way, we can implement the shaking protocols required for GKP state generation and manipulation.

State verification uses a recent proposal for direct measurement of the Wigner function of atoms in an optical lattice potential~\cite{Winkelmann2022}. This method, which was shown to work even for anharmonic traps, requires one to take advantage of the differential light shift between two hyperfine ground states of the atoms (e.g., the $\ket{F, m_F} = \ket{2,-2}$ and $\ket{1,-1}$ hyperfine states of rubidium~\cite{kuhr2011}) due to their differing vector polarizabilities. 
Similar methods of differential lattice manipulation have been used, e.g., to perform quantum walk experiments~\cite{Lam2021a}, but the drawback is the lattice wavelength must be between the D1 and D2 lines of the alkali atom used, limiting atom lifetime in the lattice.

\begin{figure}
    \centering
    \includegraphics[width=0.5\textwidth]{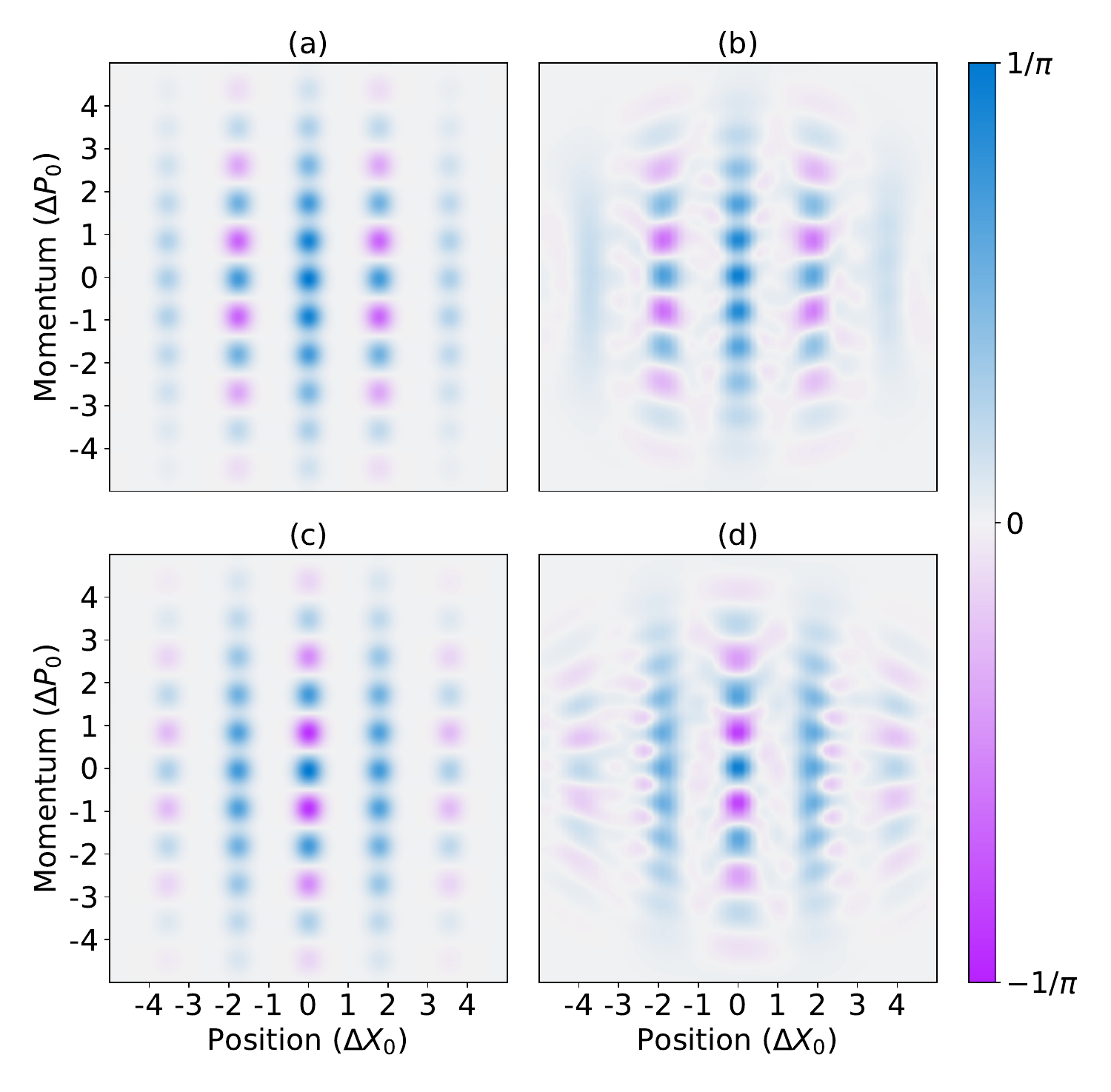}
    \caption{Wigner functions of the $\SI{10}{\deci\bel}$ squeezed $\GKP{0,1}$ states. (a) and (c) show the ideal target states (assuming $24$ bound states in the lattice) for the $\GKP{0,1}$ states, respectively. These are compared with the time-optimal $\mathcal{F}\geq 0.99$ results for $\ket{0}\rightarrow\GKP{0,1}$ states for rubidium in (b) and (d), respectively. The Wigner functions are given in units of the vacuum state variance in $x$ ($\Delta X_0$) and $p$ ($\Delta P_0$), which, for a very deep lattice, is, to an excellent approximation, the harmonic oscillator width.}
    \label{fig:WignerFunctions}
\end{figure}

It is important to note that another method for Wigner function determination via density matrix reconstruction has been demonstrated experimentally~\cite{Brown2022}. This method relies on time-of-flight data, and thus the stringent limits on lattice wavelength are relaxed. Moving to a far-off-resonant trap (e.g., at the common lattice wavelength of $\SI{1064}{\nano\meter}$) will require more lattice power to achieve a given depth (as depth goes as $I/\tilde{\Delta}$, where $I$ is the intensity of the lattice light and $\tilde{\Delta}$ is the detuning from resonance)~\cite{grimm2000}. However, the associated atom lifetime will be longer, as the photon scattering rate $\Gamma\propto I/\tilde{\Delta}^2$. We will, however, consider a lattice between the D1 and D2 lines of an alkali atom as a worst-case scenario for our proposal.

As a first step, we need to determine the lattice depth that gives us the required squeezing. From Fig.~\ref{fig:sqz-basis}, we see that to obtain $\SI{10}{\deci\bel}$ of squeezing, we require $24$ basis states and a lattice depth of $U\approx 1500 E_R$. Figure~\ref{fig:depth-lifetime} shows a contour plot of the lattice depth and scattering-limited lifetimes for Rb and Cs as a function of laser power and wavelength, restricted between the D1 and D2 lines. From the figure, we see for $U = 1500E_\R$, the best rubidium lifetime is between $5$ and $\SI{10}{\milli\second}$, but for cesium, this is about six times larger, assuming the laser power is below $\SI{100}{\milli\watt}$, which is readily achievable in the laboratory. There is a minor tradeoff, however, due to the increased wavelength of the cesium lattice that causes, in the conversion from scaled to unscaled units, the cesium protocols to be roughly a factor of two longer than the rubidium protocols, so the overall gain of a cesium system is only about three times that of rubidium. Note that for lighter alkali atoms, the achievable lattice lifetime will be lower, as the D1/D2 splitting increases with atomic mass. As such, cesium, the heaviest easily trappable alkali (although there are efforts focusing on the heavier, radioactive francium~\cite{francium}), is the best alkali atom for these experiments.

The modulation times for the time-optimal (i.e., lowest $T$ where $\mathcal{F} \geq 0.99$) protocols considered here are $\SI{141}{\micro\second} (\SI{239}{\micro\second})$ for the $\ket{0}\rightarrow\GKP{0}$ transitions in Rb (Cs) and $\SI{158}{\micro\second} (\SI{270}{\micro\second})$ for the $\ket{0}\rightarrow\GKP{1}$ transitions. The resulting rubidium Wigner functions are compared with the ideal in Fig.~\ref{fig:WignerFunctions}; Wigner results for cesium can be found in the Appendix, where we also present example controls and their spectra. Our protocols reproduce the ideal states with good fidelity; in particular, the grid-like signature of the GKP state is readily apparent. 

When considering the viability of our protocols with respect to atom lifetime in the lattice, one must also consider the time required for the Wigner measurement protocol; in Ref.~\cite{Winkelmann2022} the protocol duration was on the order of hundreds of $\SI{}{\micro\second}$; the protocol time depends on the differential trap depth, and thus, with a sufficiently large variation in the power between the different circular polarizations of the lattice light, we expect this protocol time should be maintained even at the high lattice depths required for these experiments. As such, we find, without accounting for measurement time, our modulation protocols take $\approx\SI{3}{\percent}$ of the atom lifetime in the lattice, assuming a very conservative Rb lattice lifetime of $\SI{5}{\milli\second}$. Cesium protocols require a fraction of a percent of the atom lifetime. Including an (again, conservative) $\SI{0.5}{\milli\second}$ Wigner measurement time still keeps us within $\leq\SI{15}{\percent}$ of the atom lifetime in the lattice.

Finally, we would like to remark on the scalability of this method. Empirically, we find our protocols require the lattice depth to be stable to within $\SI{0.1}{\percent}$ before substantial fidelity loss incurs. If we assume our atoms are trapped in a three-dimensional lattice with sufficiently large lattice beam waists in each dimension, the lattice depth variation across an atom cloud can be minimized. Even for modest beam waists of $\SI{150}{\micro\meter}$, one can populate $\sim 20$ lattice sites in each direction before the fidelity drops appreciably, allowing for $8000$ atoms to be simultaneously interrogated so that high-fidelity GKP states can be created along one Cartesian dimension. Additionally, state stabilization protocols can be applied along the orthogonal dimensions during state preparation so the population remains in the ground state in these dimensions; this is considered in Ref.~\cite{regalGKP} and is straightforward to include in our system as well. Thus, we expect our protocols are readily scalable to considerable atom numbers and feasible to implement with current experimental technology.

\section{\label{sec:conc}Conclusion}

In this work, we demonstrate optimal control protocols that drive an atom from the ground state of an optical lattice potential into the $\GKP{0}$ and $\GKP{1}$ states. We additionally present an experimental protocol for the generation, manipulation, and measurement of these states in optical lattices that can be readily achieved with current experimental technology. Our protocols require large lattice depths $U\approx 1500 E_R$, which poses challenges for atom lifetime and laser power, but we show that even in our worst-case scenario of rubidium atoms and lattice light between the D1 and D2 lines, the proposed protocols can be implemented within a few percent of the atom lifetime in the lattice with powers $<\SI{1}{\watt}$. We expect that as error correcting protocols improve, the required squeezing level for GKP state utility will also decrease, allowing useful atomic GKP states to be generated with lower power requirements. 
In addition, while this paper focuses on GKP state generation, extensions to gate optimization are straightforward~\cite{regalGKP} and will be the focus of future work. 

The efficient experimental generation of these states will also require finding modulation protocols that are robust, e.g., to variations in the lattice depth or noise in the modulation; this can be done via adaptation of the methods discussed in Ref.~\cite{RIM}. There is likely also utility in exploring constrained basis methods~\cite{CRAB}, for these problems, as such methods have been adapted to efficiently find state transfer protocols in other modulated-lattice systems~\cite{SLI_exp2}. These will allow for controls that can mitigate off-resonant heating effects~\cite{heating}, analogously to what was shown in~\cite{lattice_heating} for phase-modulated shallow lattices. This will also reduce the beam waist (and thus power) requirements to achieve the required lattice depths. Cavity-enhanced lattices~\cite{lattice_cavity} can also lower power requirements.

Furthermore, while our proposal centers on atoms trapped in optical lattice potentials, similar protocols can also be realized in optical tweezer potentials. Arrays of tweezer potentials do not pack atoms as tightly as lattices can, but they have the advantage that each lattice site can be more easily independently controlled, allowing for flexible continuous-variable quantum information protocols (although lattice systems with high-resolution imaging and potential projection capabilities can allow for single-site manipulation~\cite{kuhr2011}). Additionally, the proposed Wigner measurement protocol is applicable to tweezer systems.

Finally, for useful quantum information protocols to be realized, one must entangle atoms with one another. While entanglement protocols are a subject for future work, we would like to highlight that proposals for the implementation of $\sqrt{\mathrm{SWAP}}$ gates with atoms in optical lattices has been proposed in Ref.~\cite{sqrtSWAP}. Similar protocols could be used to generate cluster states of atoms~\cite{Zhou2022}, leading to a new paradigm of continuous variable quantum information in atomic systems.

In the preparation of this work, the authors became aware of similar work by Grochowski et al.~\cite{regalGKP}. Reference~\cite{regalGKP} is generally applicable to any anharmonic quantum system that admits similar driving terms to those considered here and considers a wider array of interesting quantum states, e.g., cat states~\cite{cat}. Our work, in contrast, presents a feasible experimental proposal for the generation of atomic GKP states with $\SI{10}{\deci\bel}$ of squeezing.

\section{Acknowledgments}

This work was carried out using the computational facilities of the Advanced Computing Research Centre, University of Bristol - http://www.bristol.ac.uk/acrc/. H.C.P.K. was funded by the EPSRC
CDT in quantum engineering (EP/SO23607/1). The authors would like to thank G. Ferrini and A. Alberti for interesting discussions that motivated this work and A. Clark for valuable comments on the paper draft.

\newpage
\onecolumngrid
\appendix
\section*{Appendix}

\subsection{Lifetime and Wigner results for cesium}

Here, we present plots like those of Fig.~\ref{fig:WignerFunctions} for cesium.
\begin{figure}[h]
    \centering
    \includegraphics[width=0.5\textwidth]{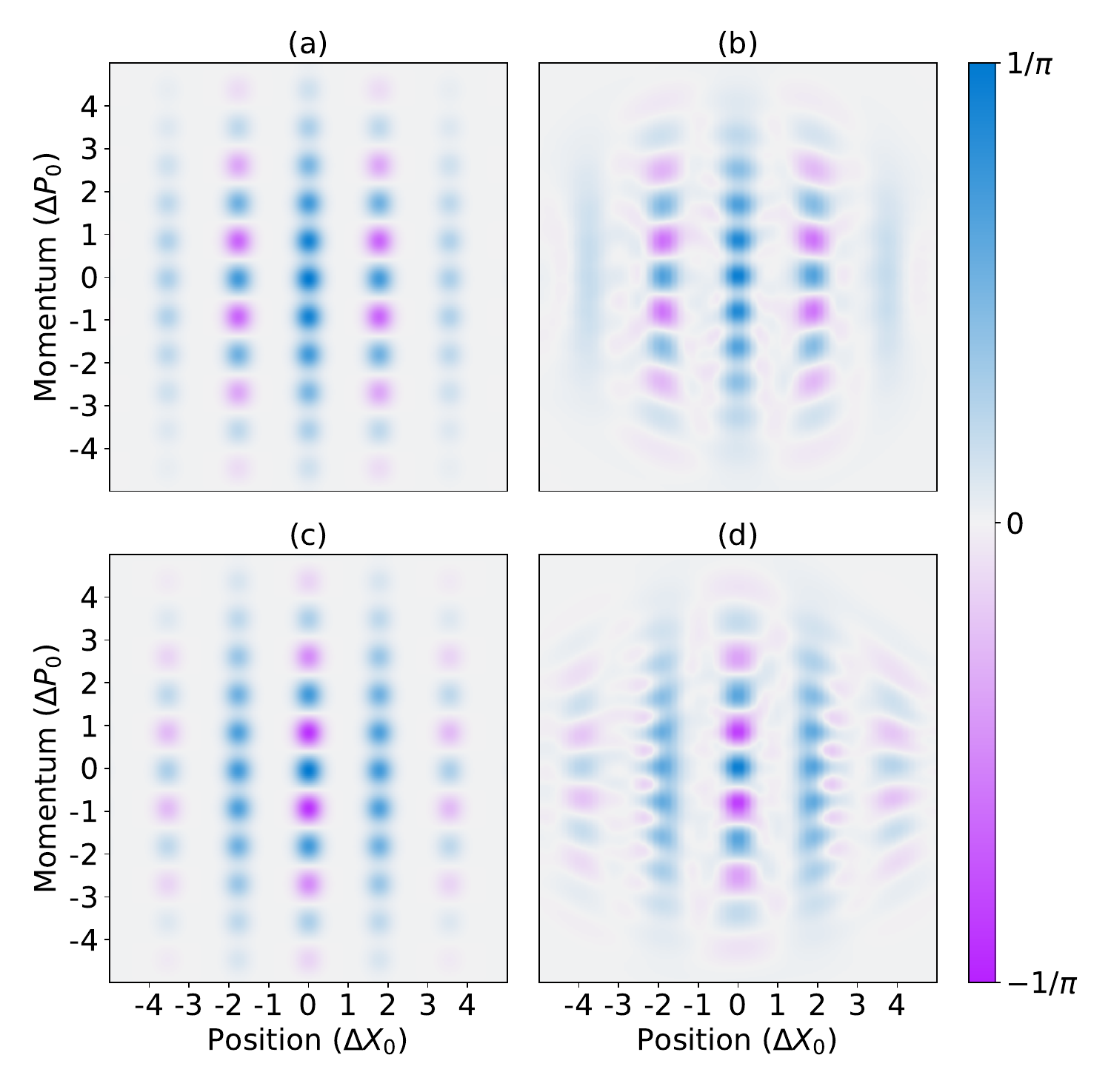}
    \caption{Same as Fig.~\ref{fig:WignerFunctions}, but for cesium in (b) and (d) for $\GKP{0,1}$ respectively. The ideal GKP state plots (a) and (c) are identical to those in Fig.~\ref{fig:WignerFunctions} for ease of comparison.}
    \label{fig:WignerFunctionsCs}
\end{figure}
\subsection{Controls and their spectra}

Here we present our controls and their Fourier spectra for the transitions considered in this work.

\begin{figure*}[h!]
    \centering
    \includegraphics[width=\textwidth]{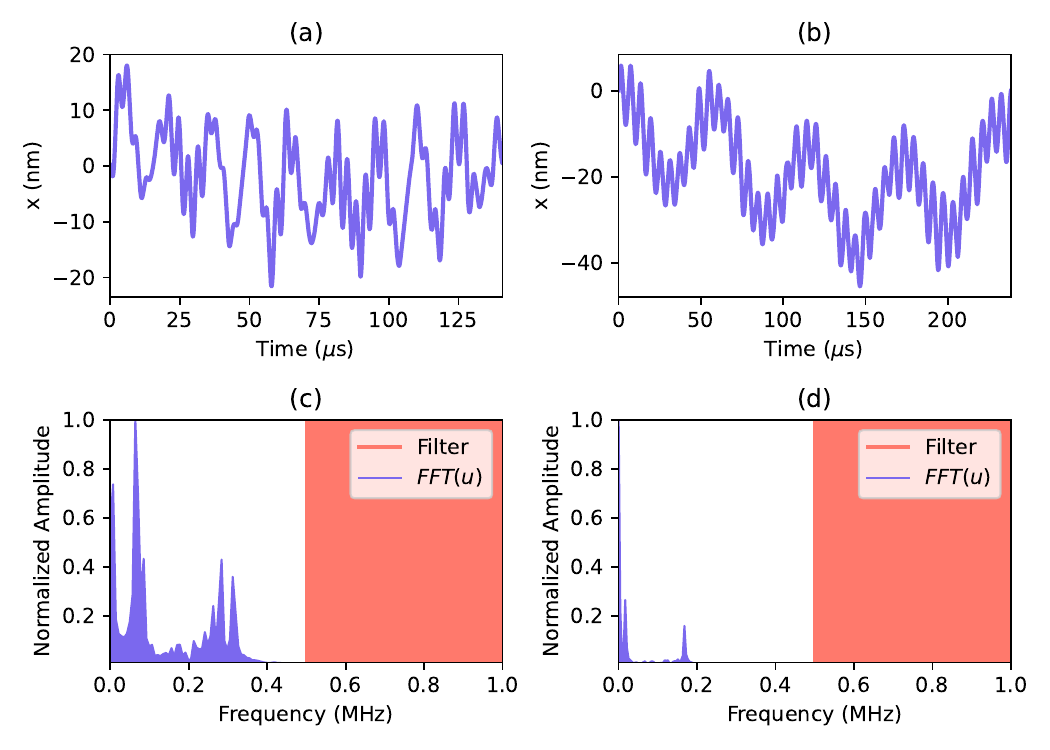}
    \caption{Control and Fourier spectrum for the time-optimal state transfer protocol driving the $\ket{0}\rightarrow\GKP{0}$ transition. The control (a), (b) and spectra (c), (d) are shown for rubidium and cesium, respectively. For (a) and (b), the blue line shows the optimized control as a function of time. In (c) and (d), the spectrum is shown in blue, and the filter applied during optimization is denoted with the excluded frequencies shaded in red.}
    \label{fig:controls0}
\end{figure*}

\begin{figure*}[h!]
    \centering
    \includegraphics[width=\textwidth]{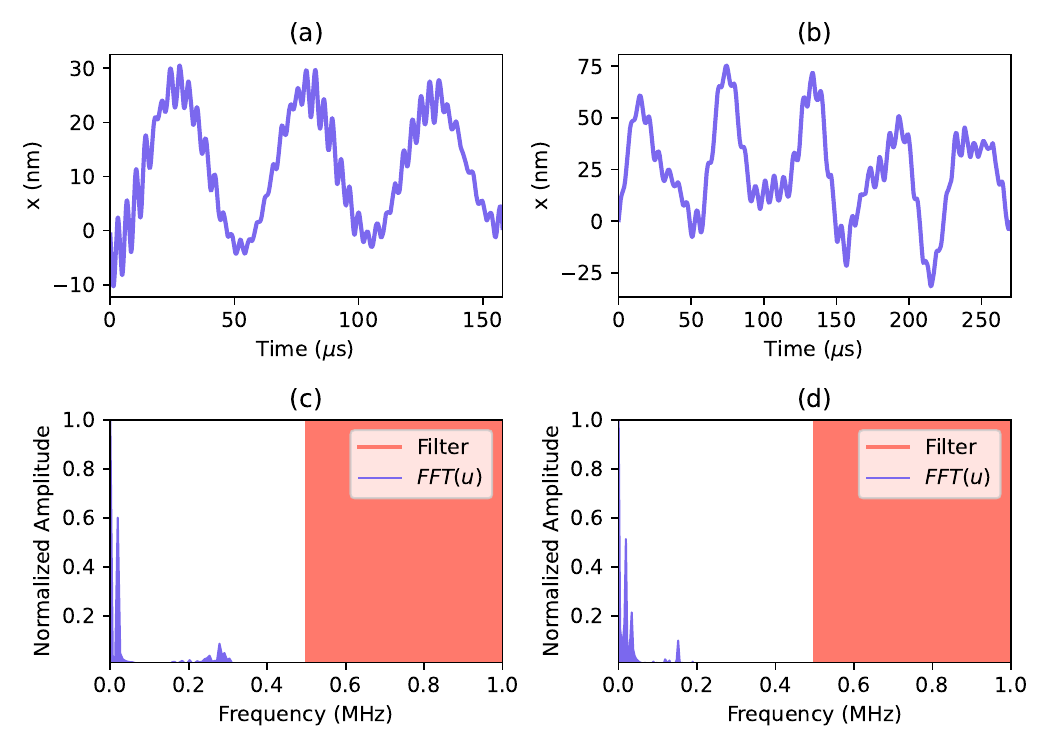}
    \caption{Same as Fig.~\ref{fig:controls0} but for the $\ket{0}\rightarrow\GKP{1}$ transition.}
    \label{fig:controls1}
\end{figure*}
\end{document}